\documentclass[twocolumn,showpacs,preprintnumbers]{revtex4}
\usepackage{amssymb}
\usepackage[dvips]{graphicx}

\begin{document}

\title{Optical conductivity of small adiabatic polarons with a long-range electron-phonon interaction}

\author{B.Ya.Yavidov$^{1,2}$}

\address{$^1$Department of Physics, Loughborough University, Leicestershire, LE11 3TU UK\\
$^{2}$Nukus State Pedagogical Institute named after A'jiniyaz,
742005 Nukus,
        Karakalpakstan, Uzbekistan}

\begin{abstract}
The hopping of an electron, interacting with many ions of a lattice via the long-range (Fr\"{o}hlich)
electron-phonon interaction and optical absorption are studied at zero temperature. Ions are assumed to be
isotropic three-dimensional oscillators. The optical conductivity and a renormalized mass of small adiabatic
Fr\"{o}hlich polarons is calculated and compared with those of small adiabatic Holstein polarons.

\end{abstract}

\maketitle

Polarons have been extensively studied since a seminal paper of Landau \cite{lan2}. They are divided into small
and large polarons in accordance with the size of their wave function. In the first case a carrier is coupled to
intramolecular vibrations and self-trapped on a single site. Its size is the same
 as the size of the phonon cloud, both are about the lattice constant (so-called small Holstein polaron (SHP)). In the case of
large polarons the size of polaron is also the same as the size of the phonon cloud, but the polaron extends
over many lattice constants. In Ref.\cite{asa-alekor} new polarons were introduced with a very different
internal structure. They were called small Fr\"{o}hlich polarons (SFP). SFP size is about the lattice constant,
but its phonon cloud spreads over the whole crystal. Within the model \cite{asa-alekor} a renormalized mass
appears to be much smaller compared with that in the canonical Holstein model \cite{hol}. Recently \cite{ay} we
extended this model to the adiabatic limit and found that the mass of SFP is much less renormalized than the
mass of SHP in this limit as well. Ref.\cite{ay} considered an electron interacting with vibrations of a chain
of ions, polarized perpendicular to the chain. However, in real systems ions vibrate in all directions. In order
to describe a more realistic case I consider here an electron hopping between two sites and interacting with
three-dimensional (3D) vibrations of nearest-neighbour ions of the chain, as shown in Fig.1(a). In addition I
calculate the optical conductivity of the system to show that the presence of an additional ion (0) (Fig.1(a))
qualitatively changes the polaron hopping and the optical conductivity compared with the Holstein model,
Fig.1(b).

\begin{figure}[tbp]
\begin{center}
\includegraphics[angle=-0,width=0.3\textwidth]{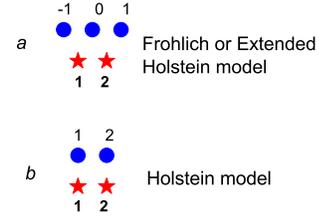} \vskip -0.5mm
\end{center}
\caption{Electron hops between sites {\textbf{1}} and {\textbf{2}}. In our model (a) electron at site {\bf 1}
interacts with 3D vibrations of $m=-1$ and $m=0$ ions, wile in the Holstein model (b) it interacts only with 3D
vibrations of $m=1$ ion.}
\end{figure}
\vspace{0.5cm}

First let us derive an analytical expression for the renormalized hopping integral of SFP in the nonadiabatic
and in the adiabatic limit in order to elucidate the effect of ion's longitudinal vibrations in the renormalized
hopping integral. The Hamiltonian of the model  is \cite{asa-alekor,hol,ay}
\begin{equation}
H=H_{ph}+H_{e}+H_{e-ph}
\end{equation}
where ($\hbar=1$)
\begin{equation}
H_{ph}=\sum_{m}\left(-\frac{\partial^2}{2M\partial {\bf
u}^{2}_m}+\frac{M\omega^2 {\bf u}^{2}_m}{2}\right)
\end{equation}
is the Hamiltonian of vibrating ions,
\begin{equation}
H_{e-ph}=\sum_{\bf i=1,2}\sum_mc^{\dagger}_{\bf i}c_{\bf i}{\bf f}_m({\bf i})\cdot{\bf u}_m
\end{equation}
describes interaction between the electron and ions, and $H_{e}=-t (c^{\dagger}_{\bf 1}c_{\bf 2} +H.c.)$  is the
electron hopping energy. Here ${\bf u}_m$ is the displacement and ${\bf f}_{m}({\bf i})$ is a force between
electron on site {\bf i} and the m-th ion. $M$ is the mass of vibrating ions and $\omega$ is their frequency.
Similar to Ref.\cite{asa-alekor,ay} the renormalized hopping integral  in the nonadiabatic case is
$\widetilde{t}=t\cdot \exp \left(-g^{2}\right)$, where
\begin{equation}
g^{2}=\frac {1}{2M\omega^3} \sum_m\left[{\bf f}^{2}_m({\bf 1})-{\bf f}_m({\bf 1})\cdot{\bf f}_m({\bf 2})
\right].
\end{equation}

One can rewrite hopping integral as $\widetilde{t}=t\cdot \exp(-\gamma E_p/\omega)$, where
$E_p=\sum_{m}(1/2M\omega^2)f^{2}_{m}(\bf 1)$ is the polaronic shift and

\begin{equation}\label{4}
    \gamma=1-\frac{\sum_{m}{\bf f}_{m}({\bf 1)}{\bf f}_{m}({\bf 2})}{\sum_{m}f^{2}_{m}(\bf 1)}.
\end{equation}

 In the nearest-neighbours approximation one can take into account only three ions in the upper chain, Fig.1(a).
Even this simplified five-site model qualitatively distinguishes from the canonical two-site Holstein model,
Fig.1(b), and maintains the features of the long-range Fr\"{o}hlich interaction. One can see that polaronic
shift and $\gamma$ factor in the Fr\"{o}hlich and the Holstein models are given by $E_p=f^{2}_{1}({\bf
1})/M\omega^2$, $\gamma=0.75$ and $E_p=f^{2}_{1}({\bf 1})/2M\omega^2$, $\gamma=1$, respectively. As a
consequence small nonadiabatic Fr\"{o}hlich polaron is less renormalized compared to small Holstein polaron with
the same $E_p$ (Fig.2). The ratio of masses of nonadiabatic SFP with three-dimensional ion vibrations to the
nonadiabatic SFP with vibrations polarized perpendicular to the chain is given by
$m_{3D}/m_{1D}=\exp(E_p/4\omega)$ for the same polaronic shift.

\vspace{0.5cm}

In the opposite adiabatic regime  we use the Born-Oppenheimer approximation representing the wave function as a
product of wave functions describing the "vibrating" ions, $\chi({\bf u}_m)$,
 and the electron with a "frozen" ion displacements, $\left(%
\begin{array}{cc}
  \psi({\bf u}_m) & \varphi({\bf u}_m) \\
\end{array}%
\right)^{T}$ ($T$ means transpose matrix). Terms with the first and second derivatives of the "electronic"
functions $\psi({\bf u}_m)$ and $\varphi({\bf u}_m)$ are small compared with the corresponding terms with
derivatives of $\chi({\bf u}_m)$. The wave function of the "frozen" state obeys the following equations

\begin{equation}
\left[E({\bf u}_m)-\sum_m{\bf f}_m({\bf 1}){\bf u}_m\right]\psi({\bf u}_m)-t\varphi({\bf u}_m)=0
\end{equation}
\begin{equation}
-t\psi({\bf u}_m)+\left[E({\bf u}_m)-\sum_m{\bf f}_m({\bf 2}){\bf u}_m\right]\varphi({\bf u}_m)=0.
\end{equation}
The lowest energy is
\begin{equation}
E({\bf u}_m)=\frac{1}{2}\sum_m{\bf f}^{+}_m {\bf u}_m-\left[\frac{1}{4}(\sum_m{\bf f}^{-}_m{\bf
u}_m)^2+t^2\right]^{1/2},
\end{equation}
that plays a role of potential energy in the equation for $\chi({\bf u}_m)$
\begin{equation}
\left[E-\sum_m H_{ph}-E({\bf u}_m)\right]\chi({\bf u}_m)=0.
\end{equation}

Here ${\bf f}^{+}_m=\left[{\bf f}_m({\bf 1})+{\bf f}_m({\bf 2})\right]$ and ${\bf f}^{-}_m=\left[{\bf f}_m({\bf
1})-{\bf f}_m({\bf 2})\right]$. By using more general transformation formulas than in Ref.\cite{ay},
\begin{widetext}
\begin{eqnarray*}
  f_{+1\alpha}^{+}u_{-1\alpha}+f_{-1\alpha}^{-}u_{+1\alpha} &=& q_{\alpha}X_{\alpha}, \\
  f_{-1\alpha}^{-}u_{-1\alpha}+(f_{+1\alpha}^{+}-f_{-1\alpha}^{-})u_{0\alpha}-f_{+1\alpha}^{+}u_{+1\alpha} &=& q_{\alpha}Y_{\alpha}, \\
\end{eqnarray*}
and
\begin{eqnarray*}
  f_{-1\alpha}^{-}(f_{+1\alpha}^{+}-f_{-1\alpha}^{-})u_{-1\alpha}-
(f_{-1\alpha}^{-2}+f_{+1\alpha}^{+2})u_{0\alpha}-f_{+1\alpha}^{+}(f_{+1\alpha}^{+}-f_{-1\alpha}^{-})u_{+1\alpha}
&=& (f_{-1\alpha}^{-2}+f_{+1\alpha}^{+2})Z_{\alpha},
\end{eqnarray*}
\end{widetext}
where $q_{\alpha}=\sqrt{2(f_{-1\alpha}^{-2}-f_{-1\alpha}^{-}f_{+1\alpha}^{+}+f_{+1\alpha}^{+2})}$, $\alpha=x,y$
and introducing a new variable $\xi=Y_{x}+Y_{y}$, one can integrate out eight of nine vibration modes and reduce
the problem to the well known double-well potential problem \cite{hol}

\begin{equation}
\left(E-4\omega+\frac{5}{8}E_p+\frac{\partial^2}{2\mu\partial\xi^{2}}-U(\xi)\right)\chi(\xi)=0.
\end{equation}
Here
\begin{equation}
U(\xi)=\frac{\mu\omega^2 \xi^{2}}{2}-\left[\frac{3}{4}\mu\omega^{2}E_{p}\xi^2 +t^2\right]^{1/2}
\end{equation}
is the familiar double-well potential, and $\mu=M/2$. Standard procedure yields for energy splitting $\Delta
E=\Delta\cdot \exp(-g^{2}_F)$, where
\begin{equation}
\Delta=\frac{\widetilde{\omega}}{\pi}\sqrt{\frac{3E_{p}}{4\omega}\kappa^{3/2}}\left(1-\sqrt{1-\left(\frac{3E_{p}}{4\omega}\kappa^{3/2}\right)^{-1}}\right),
\end{equation}
and
\begin{equation}
g^{2}_F=\frac{3E_{p}}{4\omega}\kappa^{1/2}\sqrt{1-\left(\frac{3E_{p}}{4\omega}\kappa^{1/2}\right)^{-1}}.
\end{equation}
Here $\widetilde{\omega}=\omega\sqrt{\kappa}$ is the renormalized phonon frequency, $\kappa=(1-1/36\lambda^2)$
and $\lambda=E_p/(2t)$.

The mass of adiabatic SFP with polarized (perpendicular to the chain) and three-dimensional ion vibrations is
plotted in Fig.3. The relative change of the adiabatic SFP mass $(m_{3D}-m_{1D})/m_{1D}$ is plotted as well. One
can see that the longitudinal component of ion vibrations ({\it i.e.} parallel to the chain) increases the SFP
mass compared with SHP as expected \cite{kr,trg}. Nevertheless the net contribution of all vibrations provides
much lighter adiabatic SFP than adiabatic SHP even with the 3D vibrations of ions (Fig.4).

\vspace{0.5 cm}
\begin{figure}[tbp]
\begin{center}
\includegraphics[angle=-0,width=0.47\textwidth]{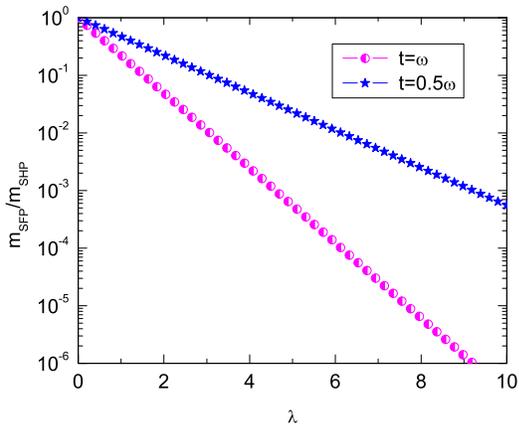} \vskip -0.5mm
\end{center}
\caption{Ratio of masses of Fr\"{o}hlich and Holstein small polarons as a function of the electron-phonon
coupling constant $\lambda$ at different values of $\omega/t$ in the nonadiabatic regime.}
\end{figure}

\vspace{0.5 cm}
\begin{figure}[tbp]
\begin{center}
\includegraphics[angle=-0,width=0.47\textwidth]{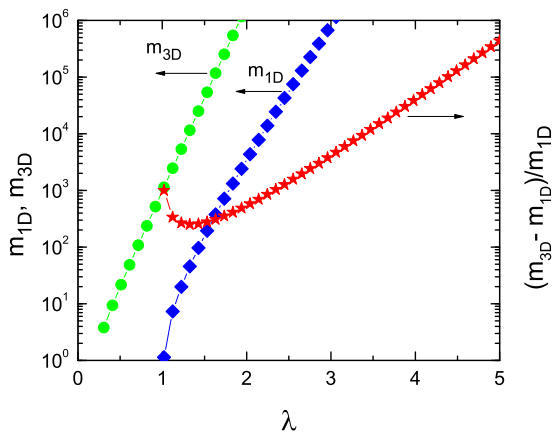} \vskip -0.5mm
\end{center}
\caption{Mass of the small Fr\"{o}hlich polaron in units of the band electron mass ($m=1/2a^{2}t$) with
polarized ($m_{1D}$) and vector ($m_{3D}$) ion vibrations, and their relative change as a function of $\lambda$
in the adiabatic regime, $t/\omega=5$.}
\end{figure}

\vspace{ 0.5 cm}
\begin{figure}[tbp]
\begin{center}
\includegraphics[angle=-0,width=0.47\textwidth]{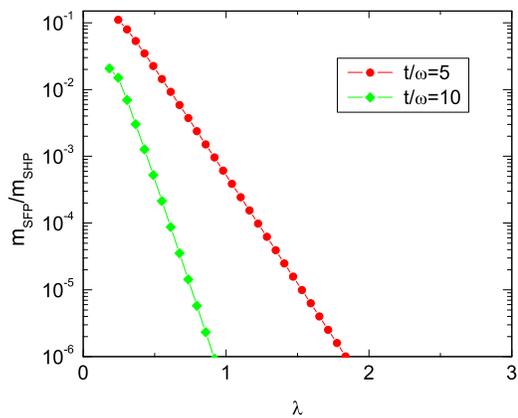} \vskip -0.5mm
\end{center}
\caption{Ratio of the SFP  mass to the SHP mass as a function of
${\lambda}$ in the adiabatic regime.}
\end{figure}
\vspace{ 0.5 cm}

Optical conductivity of both small \cite{eag,reik,klinger,bry,a-k-r} and large
\cite{devr-1,devr-2,devr-3,devr-4,devr-5,devr-6,ocf} polarons have been studied extensively. In our case of the
adiabatic small polaron the optical absorption is nearly adiabatic process so that one can apply the familiar
Franck-Condon principle. Here I  adopt a general formula for the optical conductivity of small polarons which at
$T=0$ is written as \cite{mahan}
\begin{equation}
\sigma(\nu)=\frac{\sigma_{0}\widetilde{t}^2}{\hbar\nu\sqrt{2E_{a}\hbar\omega}}exp\left[{-\frac{(\nu-4E_{a})^2}{(2\sqrt{2E_{a}\hbar\omega})^2}}\right],
\end{equation}
where $\sigma_{0}$ is a constant, $\omega$ is the phonon frequency, $\nu$ is the photon frequency and $E_{a}$ is
an activation energy for hopping process. The main difference between polarons with the Holstein and the
Fr\"{o}hlich interactions is that in the former case electron deforms only the site where it seats, while in the
second case it deforms also many neighbouring sites. This difference can be seen (i) in diagonal transitions of
polaron from site to site which ensures a lighter polaron in the Fr\"{o}hlich model and (ii) in the optical
absorption spectra. Due to the photon absorption SHP hops to an undeformed site, and $E_{a}=E_{p}/2$. However
SFP hops to a deformed neighbouring site, so that $E_{a}=\gamma E_{p}/2$. As a result, the optical
conductivities of SHP and SFP are very different, as shown in Fig.5. In our model the optical conductivity of
SFP has a more asymmetric gaussian shape. It is also different from
Ref.\cite{devr-1,devr-2,devr-3,devr-4,devr-5,devr-6,ocf}. In these works large Fr\"{o}hlich polarons were
studied using the effective mass approximation, where detailed crystal structure is irrelevant. The optical
conductivity of large polarons has an asymmetric shape with a threshold at the optical phonon frequency
$\omega$. This shape also depends on the many-body (polaron-polaron) interactions \cite{devr-6}. Depending on
approximations made the optical conductivity of large polarons could \cite{devr-2,devr-4} or could not
\cite{ocf} exhibit relaxed state peaks. While the optical conductivity of our discrete model is different, its
gross features are more reminiscent of the canonical shape of large polaron optical conductivity
\cite{devr-1,devr-2}.

\vspace{ 0.5 cm}
\begin{figure}[tbp]
\begin{center}
\includegraphics[angle=-0,width=0.5\textwidth]{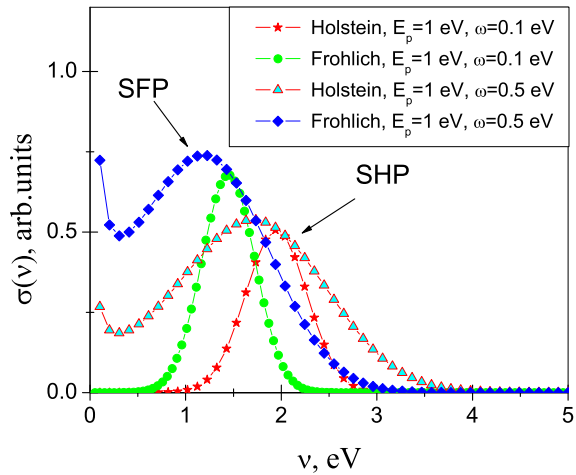} \vskip -0.5mm
\end{center}
\caption{Optical conductivity of SFP  and SHP as a function of
${\nu}$.}
\end{figure}
\vspace{ 0.5 cm}

In conclusion, I have solved an extended Holstein model with a long-range Fr\"{o}hlich interaction generalized
for the 3D vibrations. The small adiabatic Fr\"{o}hlich polaron is found many orders of magnitude lighter than
the small Holstein polaron both in the nonadiabatic (Fig.2) and adiabatic (Fig.4) regimes even with isotropic
vector vibrations of ions. The component of ions vibration parallel to the chain gives rise to a larger
enhancement of SFP mass in agrement with \cite{kr,trg}. But the common effect of all vibrations provides much
less renormalization of SFP mass compared with SHP mass. Optical conductivity of small size Fr\"{o}hlich
adiabatic polarons has been analyzed and compared with the Holstein model.

The author greatly appreciate stimulating and fruitful discussions with A.S.Alexandrov, and the financial
support of NATO and the Royal Society (grant PHJ-T3).

\end{document}